%% file: AgeDeadlineMarkovAnalysis_v5.tex
\documentclass[conference]{IEEEtran}
\usepackage{tikz}
\usepackage{acronym}
\usetikzlibrary{shapes,snakes}
\usepackage{empheq}
\usepackage{graphicx}
\usepackage{epstopdf}
\usepackage{multirow}
\usepackage{tabu}
\usepackage{array}
\usepackage{url}
\usepackage{amsmath}
\usepackage{cleveref}
\usepackage{amssymb}
\usepackage{amsthm}
\usepackage{epstopdf}
\usepackage{amsfonts}
\usepackage{tikz}
\usepackage{bm}
\usepackage{algorithmic}
\usepackage[ruled,vlined,linesnumbered]{algorithm2e}
\usepackage{graphicx}
\usetikzlibrary{arrows}
\usepackage{cite}
\usepackage{caption}
\usepackage{color}
\usepackage{subcaption}
\usepackage{amssymb}
\usepackage{tabulary}
\usepackage{booktabs}
\usepackage{soul}
\usepackage[T1]{fontenc}
\usepackage{mathtools}
\usepackage{amssymb}
\usepackage[font=small,justification=centering]{caption} 
\DeclareCaptionLabelFormat{algnonumber}{algorithm}
\definecolor{themisblue}{rgb}{0.76, 0.89, 1.0}

\tikzstyle{int}=[draw, fill=white!20, minimum size=2em]
\tikzstyle{init} = [pin edge={to-,thin,black}]

\newenvironment{list4}{
  \begin{list}{$\bullet$}{
      \setlength{\itemsep}{0.05cm}
      \setlength{\labelsep}{0.2cm}
      \setlength{\labelwidth}{0.3cm}
      \setlength{\parsep}{0in}
      \setlength{\parskip}{0in}
      \setlength{\topsep}{0in}
      \setlength{\partopsep}{0in}
      \setlength{\leftmargin}{0.17in}}}
      {\end{list}}

\input{acronyms}

\IEEEoverridecommandlockouts

\title{Information Freshness and Packet Drop Rate Interplay in a Two-User Multi-Access Channel}

\author{
Emmanouil Fountoulakis, Themistoklis Charalambous, Nikolaos Nomikos,  Anthony Ephremides, Nikolaos Pappas
	\thanks{This work has been partially supported by the Swedish Research Council (VR) and by CENIIT. The work of T. Charalambous was supported by the Academy
of Finland under Grant 317726.
		
    E. Fountoulakis and N. Pappas are with the Department of Science and Technology, Link\"{o}ping University, Campus Norrk\"{o}ping, Sweden (emails:\{emmanouil.fountoulakis, nikolaos.pappas\}@liu.se).
    
    T. Charalambous is with the School of Electrical Engineering, Aalto University, Espoo, Finland (email: themistoklis.charalambous@aalto.fi).
    
    A. Ephremides is with the Electrical and Computer Engineering Department, University of Maryland, College Park, USA (email:etony@umd.edu).
    
    N. Nomikos is with the Department of Information and Communication System Engineering, University of Aegean, Samos, Greece (email: nnomikos@aegean.gr).
	}
}
\begin{document}
	\immediate\write18{echo $PATH > tmp1}
	\immediate\write18{/Library/TeX/texbin/epstopdf > tmp2} 
	
	\maketitle
	\thispagestyle{empty}
	\pagestyle{empty}
	
	\begin{abstract}
		In this work, we combine the two notions of timely delivery of information  to study their interplay; namely, deadline-constrained packet delivery due to latency constraints and freshness of information. More specifically, we consider a two-user multiple access setup with random-access, in which user 1 is a wireless device with a queue and has external bursty traffic which is deadline-constrained, while user 2 monitors a sensor and transmits status updates to the destination. We provide analytical expressions for the throughput and drop probability of user 1, and an analytical expression for the average \ac{AoI} of user 2 monitoring the sensor. The relations reveal that there is a trade-off between the average AoI of user 2 and the drop rate of user 1: the lower the average AoI, the higher the drop rate, and vice versa. Simulations corroborate the validity of our theoretical results.
	\end{abstract}	
	\IEEEpeerreviewmaketitle 
	
	%
	%
	%
	%
	\section{Introduction}\label{sec:intro}
	
	With the proliferation of inexpensive devices with impressive sensing, computing, and control capabilities, there has been a rapid increase in Cyber-Physical Systems (CPSs) applications, such as autonomous vehicles, wireless industrial automation, environmental, and health monitoring, to name a few \cite{abd2019role, shreedhar2019age}. Such applications, however, accentuate the need for developing efficient algorithms offering timely delivery of information updates. 
	On several occasions, this requires information to arrive at the destination within a certain period (deadline-constrained) due to stringent requirements in terms of latency, while in other cases, it is required to keep the information at the destination as fresh as possible. Information timeliness or freshness at the destination is captured by a new metric, called the Age of Information (AoI) \cite{kostamonograph2017age, sunmodiano2019age}. It was first introduced in \cite{kaulyates2012real}, and it is defined as the time elapsed since the generation of the status update that was most recently received by a destination.
	
\begin{figure}[t]
	\centering
	\includegraphics[scale=0.17]{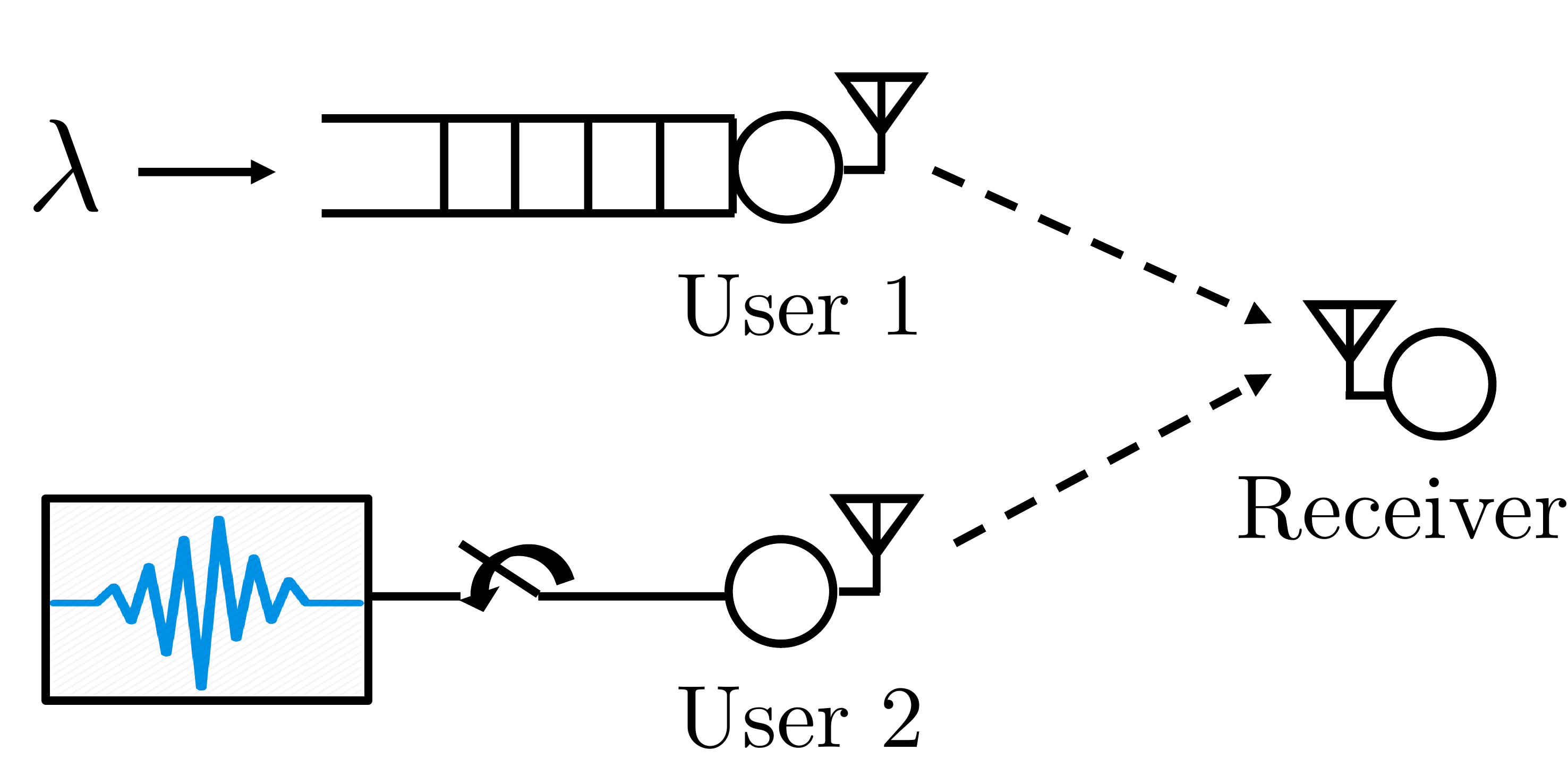}
	\caption{System model. User 1 has deadline-constrained bursty traffic, while user 2 monitors a sensor and its traffic is AoI-oriented.}
	\label{fig:sysfig}
\end{figure}

	In this work, we consider a two-user multiple access setup with different traffic characteristics: one user has external bursty traffic which is deadline-constrained, while the other user monitors a sensor that transmits status updates (in the form of packets) to the destination; this is depicted in Fig. \ref{fig:sysfig}.

	\subsection{Related Works}

	While systems with deadline-constrained traffic has been considered almost two decades ago \cite{shakkottai2002scheduling}, recently, there has been an increasing interest on studying the performance of systems with deadline-constrained traffic. For example, the works in  \cite{ModianoTWC2006, BambosTMC2007,ElAzzouni2020,HeTIT2020} consider optimal scheduling schemes for traffic with deadlines.
	The works \cite{BaeCL2013, BaeCL2015,SPAWC18}, study the performance of random access deadline-constrained wireless networks. In \cite{ManosWiOpt2017}, the authors analyze the benefits of scheduling based on exploiting variable transmission times in multi-channel wireless systems with heterogeneous traffic flows. In \cite{TepedeTVT2018},  the authors consider  a joint scheduling-and-power-allocation problem of a downlink cellular system with real-time and non-real-time users. The authors proposed an algorithm that satisfies the hard deadline requirements for the real-time users and stability constraints for the non-real-time ones. In \cite{ManosGC2018}  a dynamic algorithm that solves the problem of minimizing packet drop rate in deadline-constrained traffic by optimizing power allocation under average power consumption constraints was proposed.
	
There is a line of papers that consider the interplay of AoI with throughput or latency. For example, the work in \cite{GstamIoTJ} derives optimal status updating policies for a system with a source-destination pair that communicates via a wireless link, whereby the source node is comprised of a queue and serves two traffic flows, one that is AoI sensitive and one that is throughput oriented. 
In \cite{ZhengINFOCOM2019}, the authors study the performance of a multiple access channel with heterogeneous traffic: one grid-connected node has bursty data arrivals and another node with energy harvesting capabilities sends status updates to a common destination. The work closer to our work is \cite{SPAWC2019}, in which the interplay between delay violation probability and average AoI in a two-user wireless multiple access channel with \ac{MPR} capability  studied. Nevertheless, the authors do not consider packets with deadlines and they do not discard a packet even if the delay is larger than a threshold. On the other hand, we consider a user with packets with deadlines and therefore, we have to cope with a number of dropped packets. This changes fundamentally the problem and a different analysis approach is needed to be investigated.

\subsection{Contributions}
	
In this work, we study the interplay of deadline-constrained traffic and the information freshness in a two-user random access channel with \ac{MPR} reception capabilities. The deadline-constrained user has external bursty traffic modeled by a Bernoulli process, and the incoming packets are stored in its queue. Each packet has a predefined deadline, where if it has not been received by the destination then it is dropped from the system. The second user monitors a sensor and generates status updates at will in a timeslot. Even though this setup is small, it is tractable to analyze and it serves as a building block for more advanced setups. 
The contributions of this work are the following.
\begin{list4}
	\item For the deadline-constrained user, we provide the distribution of the waiting time of a packet and the expression for the drop rate.
	\item For the \ac{AoI}-oriented user, we provide the distribution of the \ac{AoI}, the average \ac{AoI}, and the probability the \ac{AoI} to be larger than a value for each time slot.
	\item We validate the accuracy of our analytical findings with simulations.
\end{list4}
The results show that there is a trade-off between the average AoI of user 2 and the drop rate of user 1: the lower the average AoI, the higher the drop rate, and vice versa. This is expected, since for reducing either the drop rate or the average AoI, the probability of transmission of the corresponding user should increase, causing interference to the other user.
	
\textit{To the best of our knowledge, the interaction of deadline-constrained traffic with AoI-oriented traffic has not been studied in the literature.}
	
	%
	%
	%
	%
	\section{System Model}\label{sec:model}
	We consider two users transmitting their information in form of packets over a wireless fading channel to a receiver as shown in Fig. \ref{fig:sysfig}. Time is assumed to be slotted. Let $t$ $\in$ $\mathbb{Z}_{+}$ denote the $t^{\text{th}}$ slot.

	At each time slot $t$, a packet arrives to the queue of user $1$ with arrival probability $\lambda$. Each packet $j$ of user $1$ has a deadline of $d_j$ slots since its arrival. Therefore, the packet must be successfully transmitted within $d_j$ slots; otherwise, it is dropped and discarded from the system. For simplicity of exposition, we assume that $d_j$ is the same for all packets, i.e., $d_j=d\text{, }\forall j$, which is usually the case of packets that belong to the same traffic flow. User $1$ attempts for transmission, when its queue is non-empty, with probability $q_{1}$ at each time slot. 
	
	User $2$, at each time slot, samples ``fresh" information and attempts to transmit it in form of a packet with probability $q_{2}$. We consider that the procedures of sampling together with transmission take one time slot. User $2$ discards the sampled packet after the attempted transmission.

	%
	%
	\subsection{Physical Layer Model}
	We consider that a packet from user $i$ is successfully transmitted to the receiver if and only if the signal-to-noise ratio or signal-to-interference-and-noise ratio (without or with an interferer, respectively) is above a certain threshold $\gamma_i$, i.e., $\text{SINR}_i\geq \gamma_i$. Let $P_{\text{tx},i}$  be the transmit power of user $i$, and $r_i$  be the distance between user $i$ and the receiver. The received power, when user $i$ transmits, is $P_{\text{rx},i}=h_i s_i$, where $h_i$ is a random variable (RV) representing small-scale fading and $s_i$ is the received power factor. Under Rayleigh fading, $h_i$ is exponentially distributed \cite{tse2005fundamentals}. The received power factor $s_i$ is given by $s_i = P_{\text{tx},i}r_i ^ {-\alpha}$, where $\alpha$ is the path loss exponent. When only user $i$ transmits, the success transmission probability for user $i$ is given by
$P_{i/i} = \exp \left(- \frac{\gamma_{i} \eta}{v_i s_i} \right)\text{,}$
	where $v_i$ is the parameter of the Rayleigh fading RV (i.e., $h_i\sim \mathrm{Rayleigh}(v_i)$), and $\eta$ is the noise power at the receiver. When both users transmit, the success transmission probability for user $i$ is given by \cite[Theorem~1]{nguyen2008optimization}
	 $P_{i / i, j}=\exp \left(-\frac{\gamma_{i} \eta}{v_i s_i}\right)\left(1+\gamma_{i} \frac{v_j s_j}{v_i s_i}\right)^{-1} \text{,}$
	where $j = i\mod 2 +1$.\footnote{We would like to emphasize that the analysis presented in this work is more general and it can be applied to other channel cases as long as we can obtain the values for the success probabilities.}
	Then, the service probability for user $1$  is
 	\begin{align}\nonumber 
     \mu_{1} &= q_{1} (1-q_2)P_{1/1}+q_{1}q_{2} P_{1/1,2} \\
      &=  q_1 \left[ (1-q_2)P_{1/1}  + q_2P_{1/1,2} \right]\text{,}
    \label{eq: servprob1}
 \end{align}
	and for user $2$ is 
	\begin{align}\nonumber
	\mu_{2} & = q_2(1-q_1 P\{Q>0\})P_{2/2} + q_2 q_1 (\Pr \{Q>0\} P_{2/2,1})\\
	& = q_2 \left[(1-q_1 P\{Q>0\})P_{2/2} +  q_1 (\Pr \{Q>0\} P_{2/2,1})\right]\text{,}
	\label{eq: servprob2}
	\end{align}
respectively. Then, the average success probability for user $1$ and user $2$ is
	$p_1 = (1-q_2)P_{1/1}  + q_2P_{1/1,2}\text{,}$
	and
	$p_2 = (1-q_1 P\{Q>0\})P_{2/2} +  q_1 (\Pr \{Q>0\} P_{2/2,1})\text{,}$
	respectively.

	%
	%
	%
	%
\section{Average AoI Analysis and Distribution} \label{sec:AoIanalysis}
In this section, we provide the analysis for the average and distribution of AoI. 	
\ac{AoI} represents how ``fresh" is the information from the perspective of the receiver. Let $A(t)$ be a strictly positive integer that depicts the \ac{AoI} associated with user $2$ at the receiver. The \ac{AoI} evolution at the receiver is written as
\begin{align}\label{eq:AoIev}
A(t+1) = 
\begin{cases}
1, &\text{successful packet reception,}\\
A(t)+1, &\text{otherwise.}
\end{cases}			
\end{align}
 \begin{figure}[h!]
	\centering
	\includegraphics[scale=0.38]{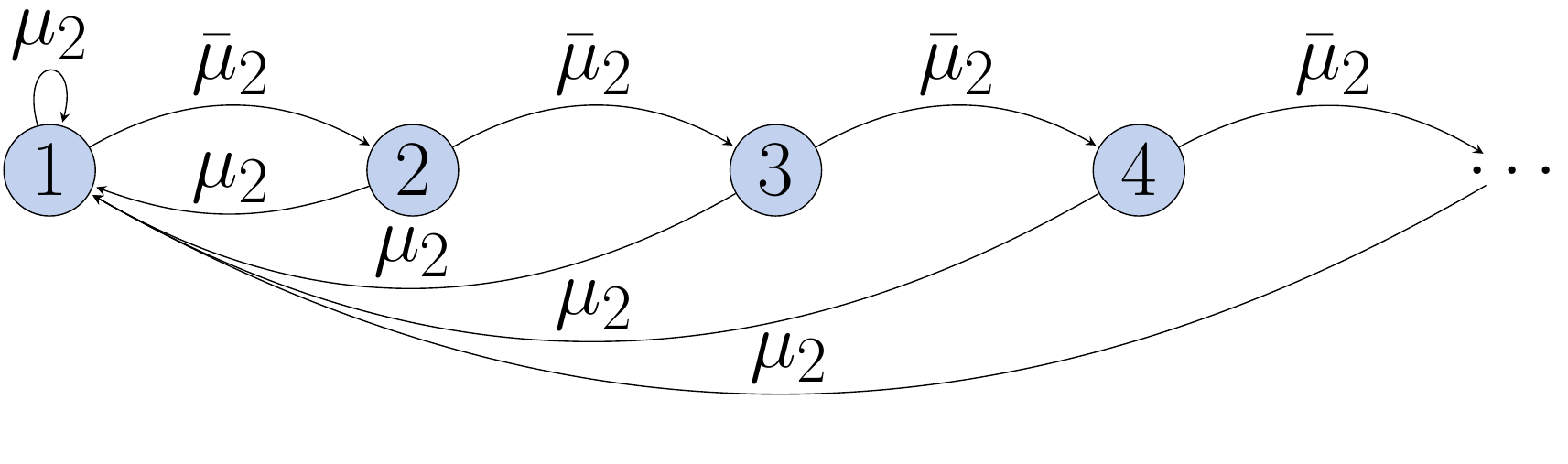}
	\caption{DTMC that models the AoI evolution.}
	\label{fig:markovianAoI}
\end{figure}
We model the evolution of the AoI as a \ac{DTMC}. According to (\ref{eq:AoIev}), at each time slot, the AoI drops to one in the case of successful packet reception from user $2$. Otherwise, it increases by one. The Markov chain is described by $P_{j\rightarrow i} = \Pr\{X_{t+1}=i\text{ }|\text{ } X_{t}=j\}\text{,}$
where $X_{t}$ denotes the value of $A(t)$ at the $t^{\text{th}}$ slot. $P_{j \rightarrow i}$ represents the 
probability to transit to state $i$ given that the current state is $j$.

The \ac{DTMC} is shown in Fig. \ref{fig:markovianAoI}, where $\bar{\mu}_{2}=1-\mu_{2}$.\footnote{For simplicity of exposition, given a probability of an event, denoted by $p$, we denote
the probability of its complementary event by 
$\bar{p}= 1 - p$.}  When the system is in state $i$, $\forall i$, it can transit only to two possible states: a) to state $1$, if we have successful packet reception; b)  to state $i+1$, otherwise. The transition matrix of the Markov chain is shown below

\vspace{0mm}
\begin{align}\label{eq:TransMatrixAge}
\begin{small}
\mathbf{P}_A=
	\begin{bmatrix}
	\mu_2 & 1-\mu_2 & 0 & 0 & \cdots \\
	\mu_2 & 0 & 1-\mu_2 & 0 & \ddots \\ 
    \vdots & \vdots & \vdots  & \ddots  & \ddots
    \end{bmatrix}\text{.}
    \end{small}
\end{align}

We denote the steady state distribution of AoI by $\bm{\pi}^A = [ \pi_{1}^A, \pi_{2}^A, \ldots]$. To obtain $\bm{\pi}^A$ we solve the following linear system of equations, $\bm{\pi}^A \mathbf{P}_A$, $\bm{\pi}^A \bm{1}=1$. Using the balance equations, we obtain
\begin{align}
 		\pi^A_i = (1-\mu_2)^{(i-1)} \mu_2\text{, }\forall i,
\end{align}
which represents the probability the \ac{AoI} to have the value of $i$.
We can now obtain the expression for the average AoI by using the steady-state distribution. The average AoI is calculated as
\begin{align}\label{eq:AveAge}\nonumber
		\bar{A} & = \sum\limits_{i=1}^{\infty} \pi_i^A i = \sum\limits_{i=1}^\infty (1-\mu_2)^{i-1}\mu_{2} i=\frac{\mu_2}{1-\mu_2} \sum\limits_{i=1}^{\infty} (1-\mu_2)^i i\\
		& \stackrel{(a)}{=}\frac{\mu_2}{1-\mu_2} \frac{1-\mu_2}{\mu_2^2} = \frac{1}{\mu_2}\text{,}
\end{align}
where $(a)$ follows by utilizing $\sum\limits_{i=1}^{\infty} ic^i = \frac{c}{(1-c)^2}\text{, } |c|<1$.

Furthermore, we calculate the probability the \ac{AoI} to be larger than a value,  $x$, where $x$ $\in$ $\mathbb{Z}_{+}$. We have that $P\left\{A>x\right\} = 1 - P\left\{A\leq x\right\} = 1-\sum\limits_{i=1}^x  \pi_{i}^A  \stackrel{(b)}{=} (1-\mu_2)^x\text{,}$
where $(b)$ follows by utilizing $\sum\limits_{i=0}^n c^i =\frac{c^{n+1}-1}{c-1}\text{, } c\neq 1$.
$P\left\{A>x\right\}$ is characterized as the \ac{AoI} violation probability which is an important metric and it indicates us the probability the \ac{AoI} to have a value larger than $x$ at each time slot.
	%
	%
	%
	%
	\section{Packet Drop Rate of User 1}\label{sec:droprate1}
	In this section, we provide the  expression for the drop rate of user $1$. We consider that if a packet from user $1$ is not successfully transmitted because of channel errors, we have the option to retransmit it. In particular, we retransmit the packet until its successful transmission or its deadline expiration. Therefore, the maximum number of retransmissions is $d-1$. 
	\begin{figure}[t!]
		\centering
		\includegraphics[scale=0.38]{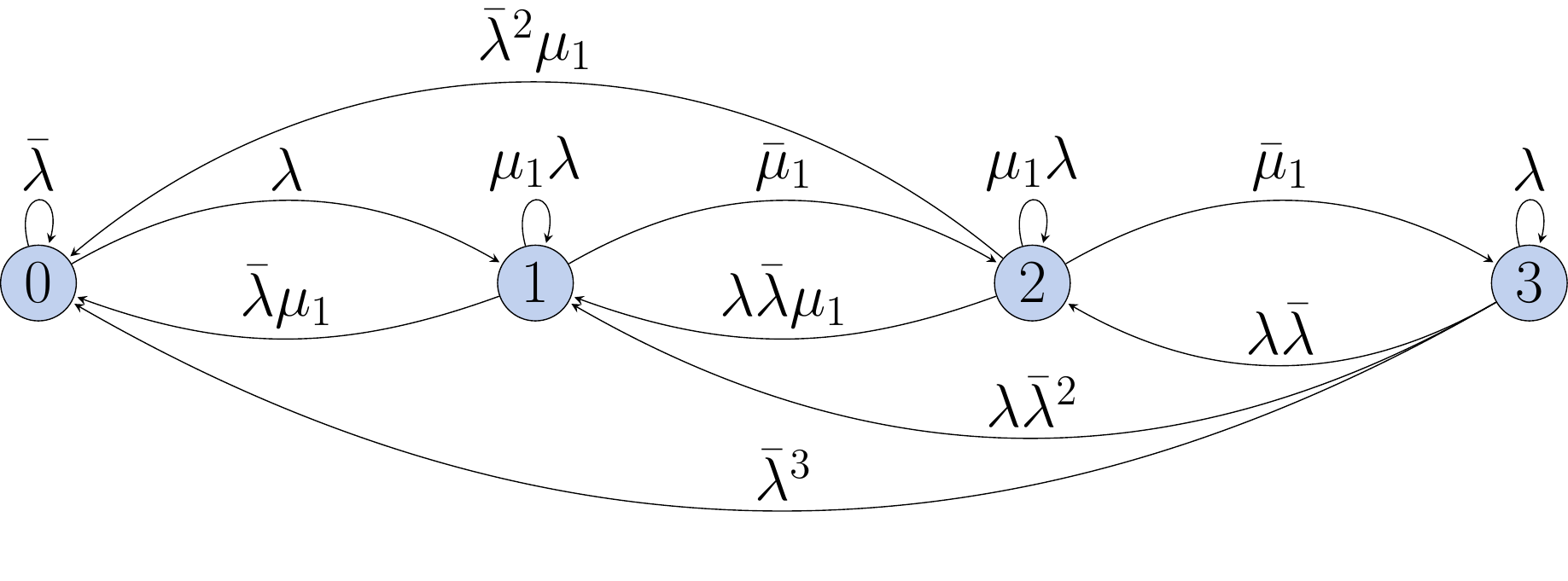}
		\caption{\ac{DTMC} for which the deadline of packets is equal to 3 time slots, i.e., $d=3$.}
		\label{fig:markovian}
	\end{figure}
	
We use a \ac{DTMC} to model the system. In particular, the states of the Markov chain represent the waiting time of the packet that is in the head of the queue. The number of states of the \ac{DTMC} is equal to $d+1$. In Fig. \ref{fig:markovian}, we depict an example of a \ac{DTMC} for a system with $d=3$. 
The system is in state $0$ if there is not packet waiting in the queue. It transits to state $1$ after the arrival of a packet. The packet experiences one slot waiting time right after its arrival because we consider the early departure - late arrival model. Therefore, the packet has the chance to be delivered in the next slot and on. When the system is in state $1$, it transits to state $0$ if the packet that is in the head of the queue has been successfully transmitted (with probability (w.p.) $\mu_1$) and no packet arrived in the current slot (w.p. $\bar{\lambda}$). From state $1$, it transits to state $2$ (w.p. $\bar{\mu}_1$) when the packet is not successfully transmitted. Finally, it remains in state $1$ when the packet that is in the head of the queue, it is successfully transmitted and a new packet arrived in the current slot (w.p. $\mu_1\lambda$).
	
The system is in state $2$ when the packet that is in the head of the queue has waiting time that equals two slots. It remains in state $2$ if the packet of the head is successfully transmitted in the current and a new packet arrived in the previous slot w.p. $\mu_1\lambda$. From state $2$, the state transits to state $1$ only if we have a successful transmission and an arrival in the current slot and no packet arrived in the previous slot (w.p. $\lambda\bar{\lambda}\mu_1$). It transits to state $0$, from state $2$, only if the packet is successfully transmitted and no packets arrived in the two previous slots (w.p. $\bar{\lambda}\bar{\lambda}\mu_1$). It transits to state $3$ only if the packet is not successfully transmitted (w.p. $\bar{\mu}_{1}$) and since the waiting time is equal to three, the packet is dropped.
	
The system remains in state $3$ only if a packet arrived three slots before the current slot (w.p. $\lambda$), therefore its waiting time equals to three slots. Since the deadline is equal to three slots, if any packet arrived at least three slots ago, either it had been transmitted or dropped. The system transits to state $1$ only if a packet arrived in the current slot and no packets arrived in the two previous slots (w.p. $\lambda\bar{\lambda}^2$). It transits to state $2$ only if a packet arrived in the previous slot and no packet arrived two slots ago (w.p. $\lambda\bar{\lambda}$). Finally, the system transits to state $0$ only if no packets arrived in the three previous slots, w.p. $\bar{\lambda^3}$.
	
The transition probability matrix (row stochastic) of the Markov chain in Fig. \ref{fig:sysfig} is shown below
	
\begin{small}
\begin{align}\nonumber
	& \mathbf{P} = \left[\begin{array}{cccc}
	\bar{\lambda}            & \lambda & 0 & 0  \\
	\mu_{1}\bar{\lambda} & \mu_{1}\lambda & \bar{\mu}_{1} & 0  \\
	\mu_{1}\bar{\lambda}^2 &  \mu_{1}\lambda\bar{\lambda}  &\mu_{1}\lambda & \bar{\mu}_1  \\
	\bar{\lambda}^3   &   \lambda\bar{\lambda}^2 & \lambda\bar{\lambda}  & \lambda
	\end{array}
	\right]\text{.}
\end{align}
\end{small}

In general, the transition matrix of the Markov chain in the general case, where the deadline is $d$, is shown below

\begin{small}
\begin{align}\nonumber
	& \mathbf{P} = \left[\begin{array}{ccccccc}
	\bar{\lambda}            & \lambda &  &  &  &  &   \\
	\mu_{1}\bar{\lambda} & \mu_{1}\lambda & \bar{\mu}_{1}&  &  &  &   \\
	\mu_{1}\bar{\lambda}^2 &  \mu_{1}\lambda\bar{\lambda}  &\mu_{1}\lambda & \bar{\mu}_1  \\
	\vdots & \vdots &\vdots& \ddots & \ddots  \\
	\mu_{1}\bar{\lambda}^{d-1} &  \mu_{1}\lambda\bar{\lambda}^{d-2} & \mu_{1}\lambda\bar{\lambda}^{d-3}  & \cdots & \mu_{1}\lambda & \bar{\mu}_1  \\
	\bar{\lambda}^d  &   \lambda\bar{\lambda}^{d-1} & \lambda\bar{\lambda}^{d-2}  & \cdots &  \bar{\lambda} & \lambda
	\end{array}
	\right]\text{.}
\end{align}
\end{small}
\vspace{0mm}
We denote by $\bm{\pi} = [\pi_{0} \text{ }\pi_{1}\text{ } \ldots\text{ } \pi_{d} ]$ the steady-state distribution of the Markov chain. To derive $\bm{\pi}$, we solve the following linear system  of equations  
$\bm{\pi}\mathbf{P}= \bm{\pi}\text{, } \bm{\pi 1} =1\text{.}$
We observe that $\bm{\pi}$ is an eigenvector of $\mathbf{P}$.
After applying  eigenvalue decomposition  we obtain the eigenvectors and eigenvalues of matrix $\mathbf{P}$. We find the eigenvector that corresponds to the eigenvalue that is equal to $1$. We normalize the elements of the eigenvector and we obtain $\bm{\pi}$. Then, we calculate the drop rate as $\bar{D}  =  \pi_{d}\bar{\mu}_1\text{.}$
In addition, we  calculate the probability the queue of user $1$ to be non empty; $\Pr\{Q>0\} = 1 - \pi_0.$ Therefore, all the terms in $\eqref{eq:AveAge}$ are now known and the average \ac{AoI} can be computed.
\subsection{Discussion on the lumpability of DTMC}
\ac{DTMC} in Fig. \ref{fig:markovian} can be an aggregated form of a 2D \ac{DTMC} which takes into account the actions of user $1$ as individual cases. The 2D \ac{DTMC} is described by $P^\text{2D}_{(i,j)\rightarrow (u,k)} =\operatorname{Pr}\left\{X_{t+1}=u, Y_{n+1}=k \mid X_{n}=i, Y_{n}=j\right\}$, where $X_t$ and $Y_t$ denote the states of the action of user $2$ and the waiting time in the queue for a packet of user $1$, respectively. Note that $X_t$ can take either  the value of one (if user $2$ is active) or zero (if user $2$ is silent).
Note that 
$P_{(0,i) \rightarrow (0,j)  }  = P_{(1,i)\rightarrow (0,j)} $ and  $P_{(0,i) \rightarrow (1,j)  }  = P_{(1,i)\rightarrow (1,j)} $, $\forall i, j$, because the action of user $2$ in the previous slot does not affect the transition in the current slot. Therefore, 
$P_{(0,i) \rightarrow (0,j)  } +P_{(0,i) \rightarrow (1,j)  }  = P_{(1,i)\rightarrow (0,j)} +P_{(1,i)\rightarrow (1,j)} $, $\forall i,j$.  Let us consider a partition of the states that is defined as $\mathcal{A}_j = \left\{ (0,j), (1,j) \right\}$. According to Theorem 6.3.2  \cite[Theorem 6.3.2, Chapter 6]{kemeny1976} the Markov Chain is lumpable with respect to the partition $\mathcal{A} = \left\{\mathcal{A}_1, \mathcal{A}_2, \ldots, \mathcal{A}_d\right\}$. Therefore, \ac{DTMC} in Fig. \ref{fig:markovian} is an equivalent Markov chain with the 2D Markov chain described above.

	%
	%
	%
	%

\section{Simulation and Numerical Results}\label{sec:simulations}
In this section, we provide results that show the interplay between the packet drop rate of user $1$ and average AoI of user $2$ at the receiver. Also, we validate our analysis by comparing the analytical with the simulation results.  
\begin{figure}[t!]
	\centering
	\includegraphics[scale=0.38]{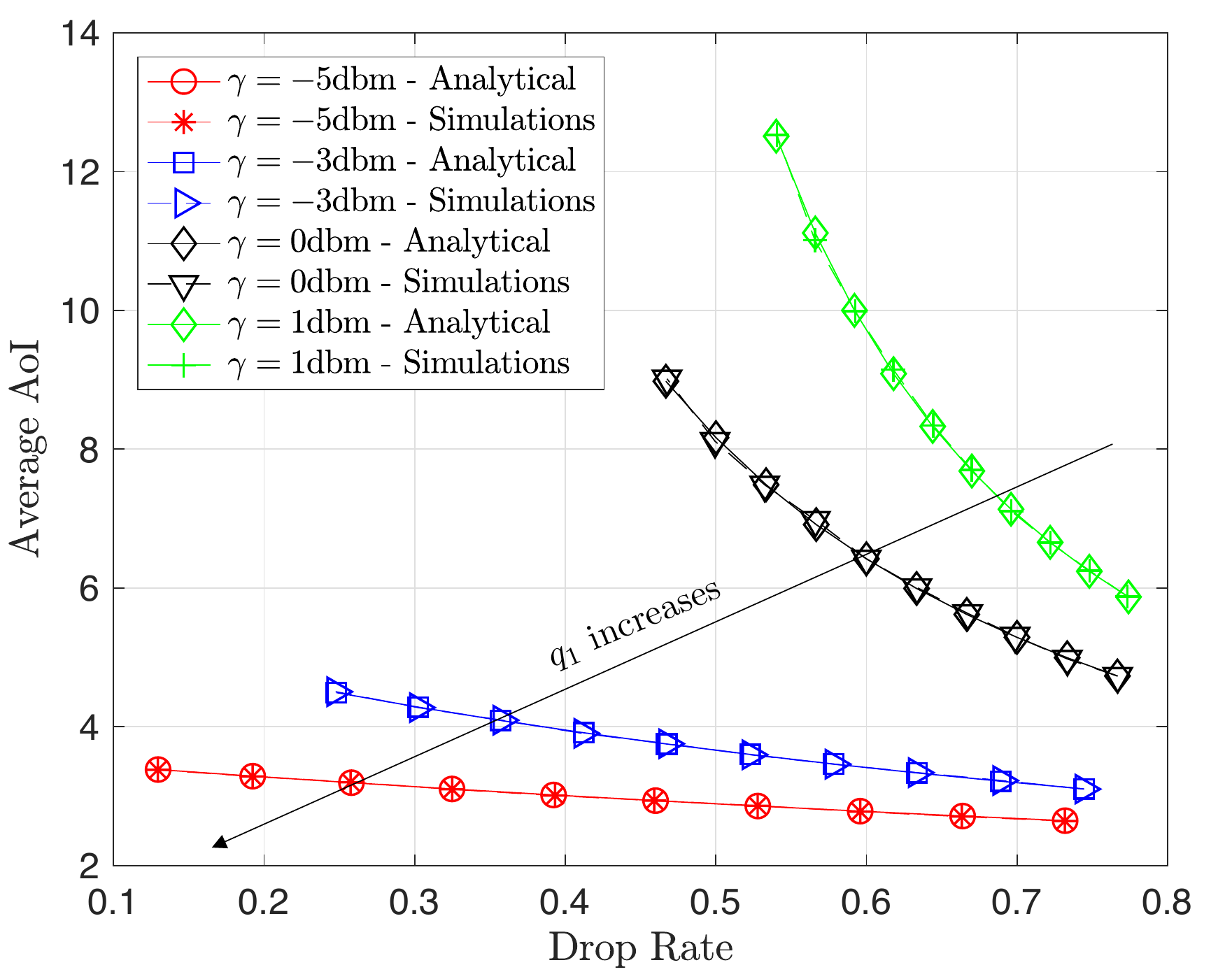}
	\caption{$q_2=0.5$, $\lambda=0.8$. $q=0.1, 0.2, \ldots, 1$.}
	\label{fig:aoidropratediffgammaq1}
\end{figure}
\begin{figure}[t!]
	\centering
	\includegraphics[scale=0.38]{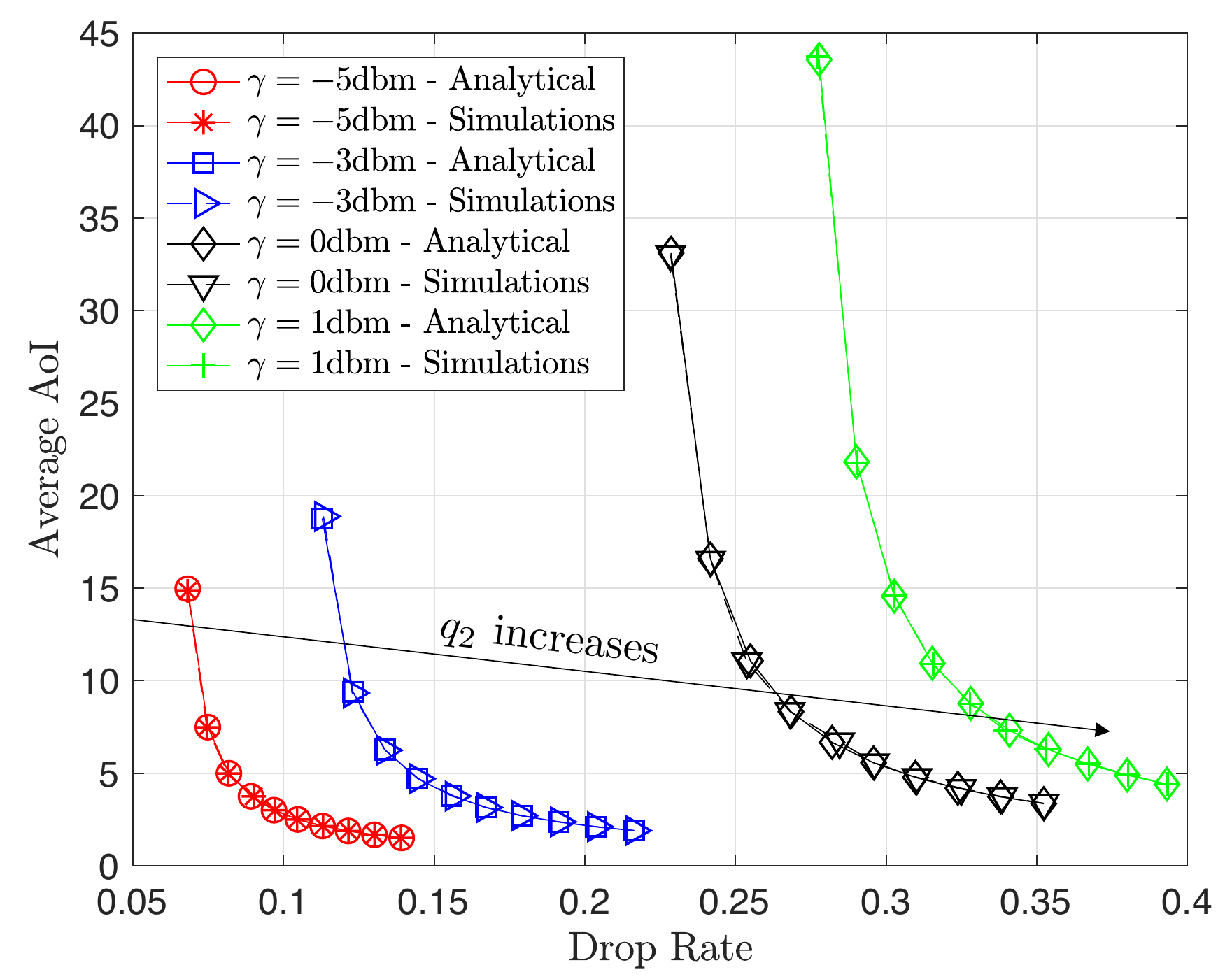}
	\caption{$q_1=0.5$, $\lambda=0.5$. $q_2=0.1, 0.2, \ldots, 1$.}
	\label{fig:aoidropratediffgammaq2}
\end{figure}
We consider  different scenarios. However, at each scenario we consider that the users are located at distance $r_i=30$m from the receiver. The receiver noise power is $\eta=-100$dbm, and the path loss exponent is $\alpha=4$. Also, both users transmit with power $P_1=P_2=5$ mW.

In Fig. \ref{fig:aoidropratediffgammaq1}, and Fig. \ref{fig:aoidropratediffgammaq2}, we depict the interplay between the average AoI of user $2$ and packet drop rate of user $1$. In Fig. \ref{fig:aoidropratediffgammaq1}, we show the effect of the value of $q_1$ on the average AoI. We consider four different cases each one with different \ac{MPR} capabilities. We denote that a receiver has strong \ac{MPR} capabilities only if $\delta=\frac{P_{1/1}}{P_{1/1,2}}  +  \frac{P_{2/2}}{P_{2/2,1}} >1$, otherwise we consider that the receiver has weak MPR capabilities. For $\gamma=-5$dbm and $-3$dbm, $\delta=1.5195$ and $1.3323$, respectively. Therefore, the receiver has weak  MPR capabilities. For $\gamma=0$dbm and $1$, $\delta=1 $ and   $0.8854$, respectively. Therefore, the receiver has weak MPR capabilities. 

In Fig. \ref{fig:aoidropratediffgammaq1}, we consider that the sampling probability is $q_2=0.5$. We obtain the drop rate and average AoI for different values of $q_1$. We observe that when the receiver has strong \ac{MPR} capabilities, the access probability of user $1$ does not significantly affect the average AoI (red line). Therefore, in this case, we obtain that the best strategy would be to allow both users to transmit at the same time. On the other hand, we observe that as the \ac{MPR} capabilities become weak the access probability of user $1$ significantly affects the average AoI (black and green lines).

In Fig. \ref{fig:aoidropratediffgammaq2}, we consider that the access probability of user $1$, $q_1$, is fixed and equal to $0.5$. Also, the arrival rate is $\lambda=0.5$. In this scenario, we consider different values of the sampling probability. We observe that when the receiver has strong \ac{MPR} capabilities (red line), we can significantly decrease the average AoI while keeping the drop rate low for user $1$ (red line). However, as the \ac{MPR} capabilities become weaker the drop rate is affected by higher values of $q_2$. To give a more realistic example, let us consider that our goal is to keep the average AoI below $5$. For $\gamma=-5$ dbm, we observe that we can achieve this target for sampling rate $q_2=0.3$ and drop rate is $0.17$. Thus, allowing both users to transmit is beneficial. For the case which $\gamma=1$dbm (weak \ac{MPR} capabilities), we observe that, in order to keep the average AoI below $5$, we should increase the sampling probability $q_2$ to the value of 
 $0.7$. However, the drop rate for user $1$ is high and it is equal to $0.41$, i.e., almost the half of the packets are dropped. In this case, a time sharing scheme will be more beneficial.
 
\begin{figure}[t!]
	\centering
	\includegraphics[scale=0.38]{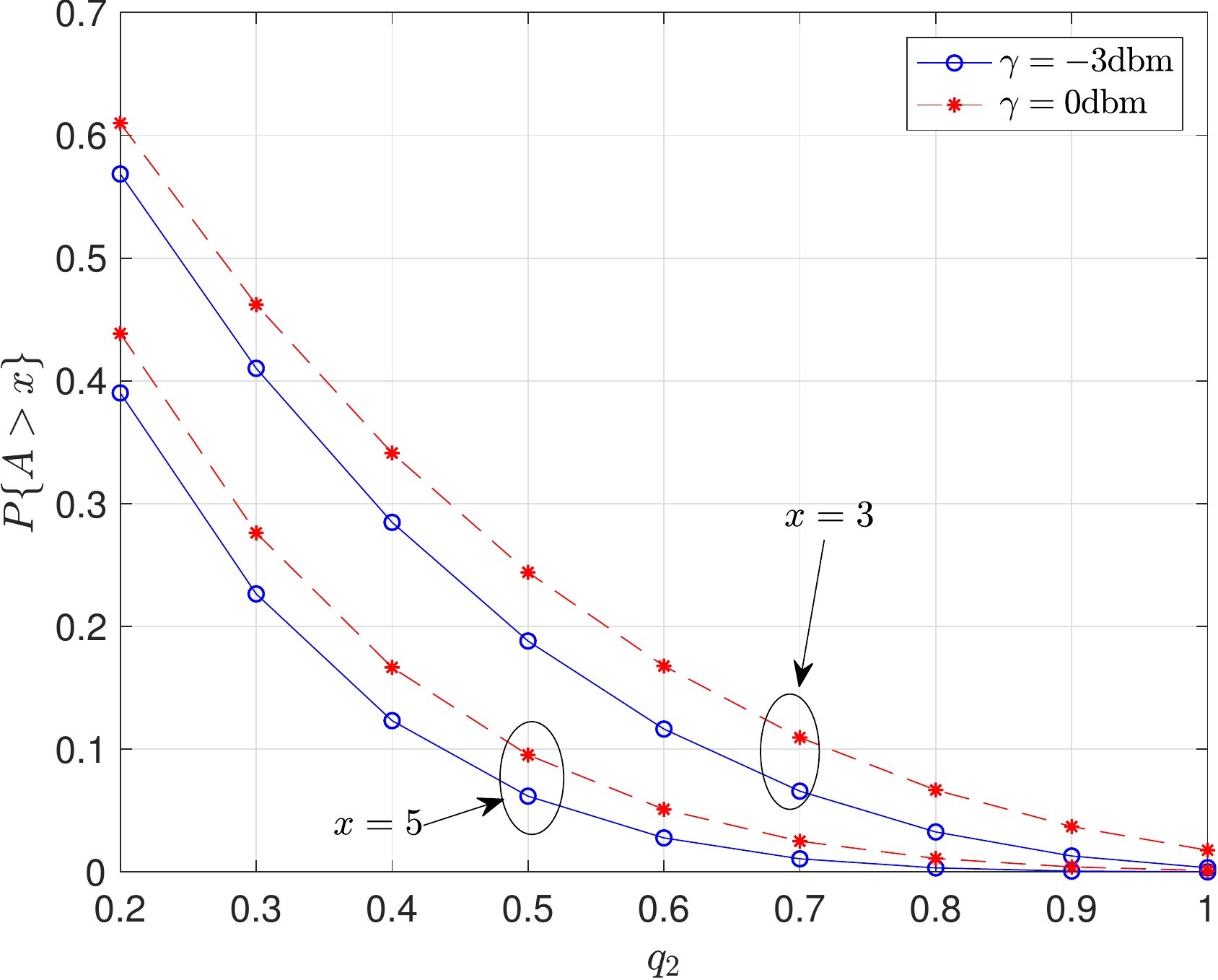}
	\caption{Probability the AoI of user $2$ at the receiver to be larger than a value, $x$. $\lambda=0.5$, $q_1=0.5$. Strong ($\gamma-3$dbm) and weak ($\gamma=0$dbm) MPR capabilities.}
	\label{fig:probdiffgamma}
\end{figure}

In Fig. \ref{fig:probdiffgamma}, we show how the value $P\left\{A>x\right\}$ changes for different values of $q_2$ and $\gamma$. As we increase the value of $q_2$ the probability decreases because user $2$ samples and attempts for transmission more often. However, due to the error-prone channel, we observe that $P\left\{A>x\right\}$ has higher values for the case of weak \ac{MPR} capabilities. Note that this metrix, i.e., $P\left\{A>x\right\}$, is important when we study the  AoI slot by slot. More precisely, based on this metric, we know what is the probability the AoI to be greater than a value of $x$ at a time slot. Therefore, we can configure the system, e.g., the value of $q_2$, based on our targets.
	
	%
	%
	%
	%
	\section{Conclusions and Future Directions}\label{sec:conclusions}
	
	In this work, we studied the interplay of deadline-constrained packet delivery and freshness of information at the destination. More specifically, we considered a two-user multiple access setup with random access, in which user 1 is a wireless device with a queue and has external bursty traffic which is deadline-constrained, while user 2 monitors a sensor and transmits status updates to the destination. We provided analytical expressions for the throughput and drop probability of user 1, and an analytical expression for the average \ac{AoI} of user 2. We demonstrated that there exists a trade-off between the average AoI of user 2 and the drop rate of user 1.  Our analytical findings are validated through simulations.
	
	From our results it is evident that the probability of accessing the channel affects the performance of individual users as well as that of the overall system. Ongoing work focuses on optimizing these probabilities. Furthermore, larger and more general setups will be considered.
	
	\bibliography{MyBibAge1}
	\bibliographystyle{ieeetr}
\end{document}

%% file: acronyms.tex
\acrodef{AoI}{Age of Information}
\acrodef{IoT}{Internet of Things}
\acrodef{DPP}{Drift-Plus-Penalty}
\acrodef{DTMC}{Discrete-Time Markov Chain}
\acrodef{MPR}{Multi-Packet-Reception}